# Lattice QCD at finite temperature : present status

Péter Petreczky[a,b]

[a]RIKEN-BNL and Physics Department, Brookhaven National Laboratory, Upton, NY, 11973, USA
[b]Kavli Institute for Theoretical Physics China, CAS, Beijing 100190, China

**Abstract**

I review recent progress in finite temperature lattice calculations, including calculations of Equation of State, fluctuations of conserved charges and spatial correlation functions. I compare lattice results with the predictions of hadron resonance gas model, resummed perturbation theory and 3-dimensional effective field theory. Comparison of the lattice results for certain ratios with the prediction of AdS/CFT correspondence is also discussed.

## 1. Introduction

Attempts to study QCD thermodynamics on the lattice go back to the early 80's when lattice calculations in $SU(2)$ gauge theory provided the first rigorous theoretical evidence for deconfinement [1]. Lattice calculations of QCD thermodynamics with light dynamical quarks remained challenging until recently. During the past 5 years calculations with light $u, d$ quarks have been performed using improved staggered fermion actions [2-9]. At zero temperature calculations with improved staggered fermions successfully reproduced many experimentally observed quantities [9, 10, 11] and even made predictions, which later were confirmed by experiment [12, 13].

To get reliable predictions from lattice QCD the lattice spacing $a$ should be sufficiently small relative to the typical QCD scale, i.e. $\Lambda_{QCD} a \ll 1$. For staggered fermions discretization errors go like $O((a\Lambda_{QCD})^2)$ but discretization errors due to flavor symmetry breaking turn out to be numerically quite large. To reduce these errors one has to use improved staggered fermion actions with so-called fat links [14]. At high temperature the dominant discretization errors go like $(aT)^2$ and therefore could be very large. Thus it is mandatory to use improved discretization schemes, which improve the quark dispersion relation and eliminate these discretization errors. Lattice fermion actions used in numerical calculations typically implement some version of fat links as well as improvement of quark dispersion relation and are referred to as $p4$, *asqtad* and *stout*. In lattice calculations the temperature is varied by varying the lattice spacing at fixed value of the temporal extent $N_\tau$. The temperature $T$ is related to lattice spacing and temporal extent, $T = 1/(N_\tau a)$. Therefore taking the continuum limit corresponds to $N_\tau \to \infty$ at the fixed physical volume.

In this contribution I am going to review lattice calculations of the equation of state, fluctuations of conserved charges and spatial correlation functions. In the past years attempts to calculate different spectral functions, in particular, quarkonium spectral functions [15, 16, 17, 18] as well as spectral functions of energy momentum tensor [19] have been done. These topics will be discussed in contributions by Mócsy [20] and Meyer [21].

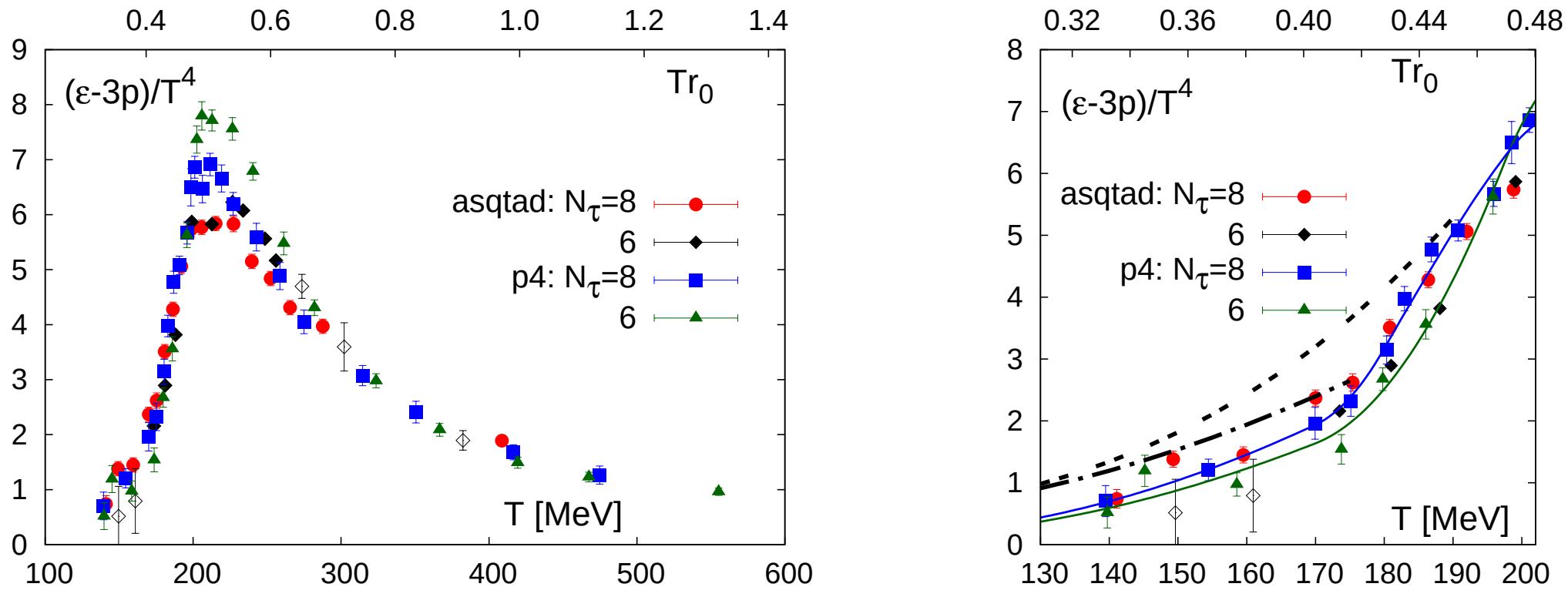

Figure 1: The interaction measure calculated with *p*4 and *asqtad* actions in entire temperature range (left) and at low temperatures (right) from Ref. [8]. The dashed and dashed-dotted lines are the prediction of hadron resonance gas (HRG) with all resonances included up 2.5GeV (dashed) and 1.5GeV (dashed-dotted), respectively.

## 2. Equation of State

The equation of state has been calculated with *p*4 and *asqtad* action on lattices with temporal extent $N_\tau = 4,\ 6$ and 8 [6, 7, 8]. In these calculations the strange quark mass was fixed to its physical value, while the light $(u,\ d)$ quark masses 10 times smaller than the strange quark mass have been used. These correspond to pion masses of $(220-260)$ MeV. The calculation of thermodynamic observables proceeds through the calculation of the trace of the energy momentum tensor $\epsilon - 3p$ also known as trace anomaly or interaction measure. This is due to the fact that this quantity can be expressed in terms of expectation values of local gluonic and fermionic operators. The explicit expression for $\epsilon - 3p$ in terms of these operators for *p*4 and *asqtad* actions can be found in Ref. [8]. Different thermodynamic observables can be obtained from the interaction measure through integration. The pressure can be written as

$$\frac{p(T)}{T^4} - \frac{p(T_0)}{T_0^4} = \int_{T_0}^{T} \frac{dT'}{T'^5}(\epsilon - 3p). \tag{1}$$

The lower integration limit $T_0$ is chosen such that the pressure is exponentially small there. Furthermore, the entropy density can be written as $s = (\epsilon + p)/T$. Since the interaction measure is the basic thermodynamic observable in the lattice calculations it is worth discussing its properties more in detail. In Fig. 1 I show the interaction measure for *p*4 and *asqtad* actions for two different lattice spacings corresponding to $N_\tau = 6$ and 8. In the high temperature region, $T > 250$ MeV results obtained with two different lattice spacings and two different actions agree quite well with each other. Discretization errors are visible in the temperature region, where $\epsilon - 3p$ is close to its maximum as well as in the low temperature region. At low temperatures the lattice data have been compared with the hadron resonance gas (HRG). As one can see the lattice data fall below the resonance gas value. This is partly due to the fact that the light quark masses are still about two times larger than the physical value as well as to discretization errors. The $O((a\Lambda_{QCD})^2)$ discretization errors in the hadron spectrum are suppressed at high temperatures as the lattice spacing $a$ is small there. Also hadrons are not the relevant degrees of freedom in this temperature region. But at small temperatures, where hadrons are the relevant degrees of freedom, these discretization effects maybe significant.

The pressure, the energy density and the speed of sound are shown in Fig. 2. The energy density shows a rapid rise in the temperature region $(185 - 195)$ MeV and quickly approaches about 90% of the ideal gas value. The pressure rises less rapidly but at the highest temperature it is also only about 15% below the ideal gas value. In the previous calculations with the $p4$ action it was found that the pressure and energy density are below the ideal gas value by about 25% at high temperatures [22]. Possible reason for this larger deviation could be the fact that the quark masses used in this calculation were fixed in units of temperature instead being tuned to give constant meson masses as lattice spacing is decreased. As discussed in Ref. [23] this could reduce the pressure by $10 - 15\%$ at high temperatures. The speed of sound reaches its smallest value at energy densities $\epsilon \simeq 1$ GeV/fm$^3$ (softest point) but it is also smaller than the conformal value $1/\sqrt{3}$ for a large range of energy densities relevant for RHIC.

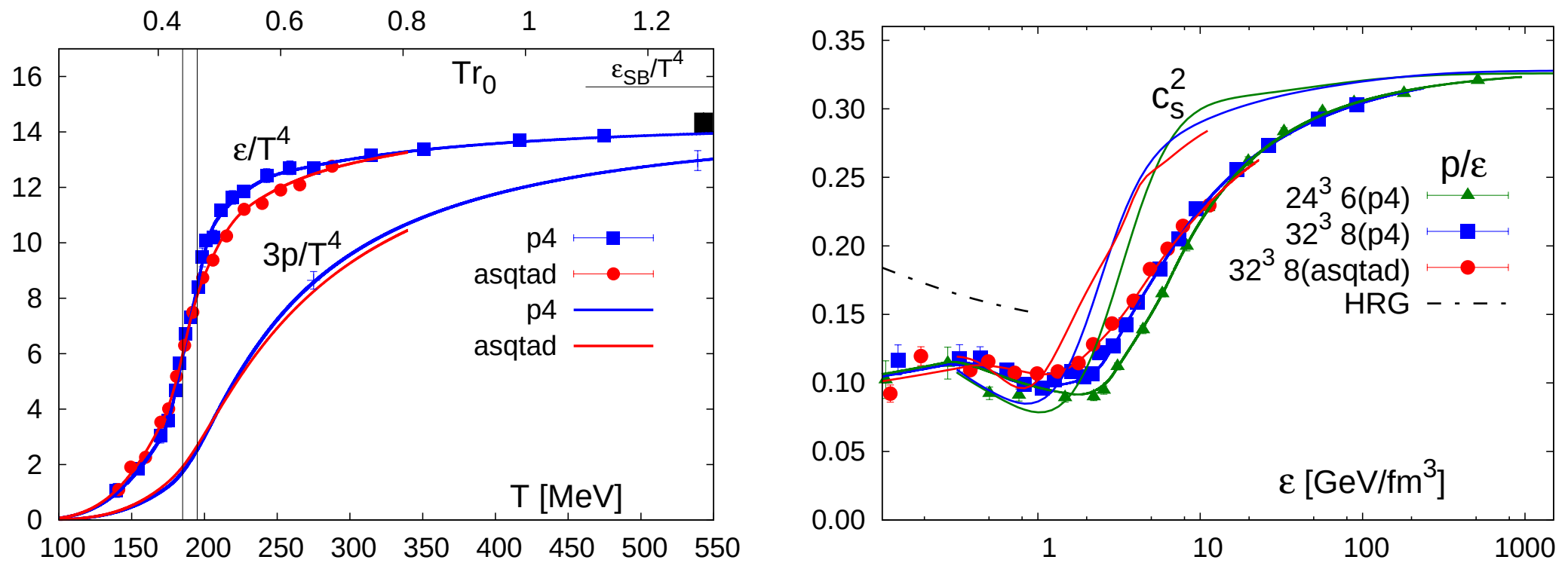

Figure 2: The energy density and the pressure as function of the temperature (left), and the the speed of sound squared and $p/\epsilon$ as function of the energy density (right) calculated with $p4$ and *asqtad* actions on $N_\tau = 8$ lattices [8]. In the left figure the vertical band denotes the transition region $T = (185 - 195)$ MeV, while the horizontal box is the uncertainty from the lower integration limit (see Ref. [8] for details).

Since at sufficiently high temperatures QCD is expected to be weakly coupled it is interesting to compare lattice results on the trace anomaly with the results obtained via weak coupling techniques. In Fig. 3 I show the trace anomaly at high temperatures only and compare it with leading order perturbative result $3b_0 g^4(\mu)T^4/4$ for two choices of the renormalization scale $\mu = \pi T$ and $\mu = 2\pi T$. Here $b_0$ is the coefficient of the 1-loop beta function. The leading order perturbative prediction overlaps with lattice data only at the highest temperature. Because the trace anomaly receives contribution from the non-perturbative gluon condensate $\langle F_{\mu\nu}^2 \rangle \sim \Lambda_{QCD}^4$ deviations from the perturbative results are in fact expected unless $T \gg \Lambda_{QCD}$. It turns out, however, that it is possible to fit the lattice data on the trace anomaly with a simple form which combines the leading order perturbative result and the expected non-perturbative correction: $\epsilon - 3p = 4B + 3b_0 g^4(\mu = \pi T)T^4/4$. The fit gives $B \simeq (200\ \mathrm{MeV})^4$ and is shown in Fig. 3. Let me finally mention that the trace anomaly has been calculated in the effective theory approach called EQCD [24]. In this approach hard modes are treated perturbatively leading to a 3-dimensional effective field theory which then can be solved on the lattice [25].

The non-perturbative constant $B$ which is present in the pressure and energy density is absent in the entropy density $s$, which can be written as $s = \partial p/\partial T$. Therefore the entropy density is the most suitable quantity for comparison with the high temperature perturbation theory. It is well know that the perturbative series for pressure in terms of powers of $g$ has very bad con-vergence properties. The convergence can be greatly improved by using resummed perturbative approaches [26, 27, 28]. The entropy density is shown in Fig. 3 together with the results of the perturbative calculations. As one can see there is a very good agreement between resummed perturbative results and the lattice data. In the figure the results from EQCD are also shown. These calculations too agree well with the lattice. The entropy density has been calculated in $N = 4$ SUSY using AdS/CFT correspondence in the limit of infinite t'Hooft coupling and was found to be 3/4 of the ideal gas value [30]. This result is also show in Fig. 3 (right) as a solid black line.

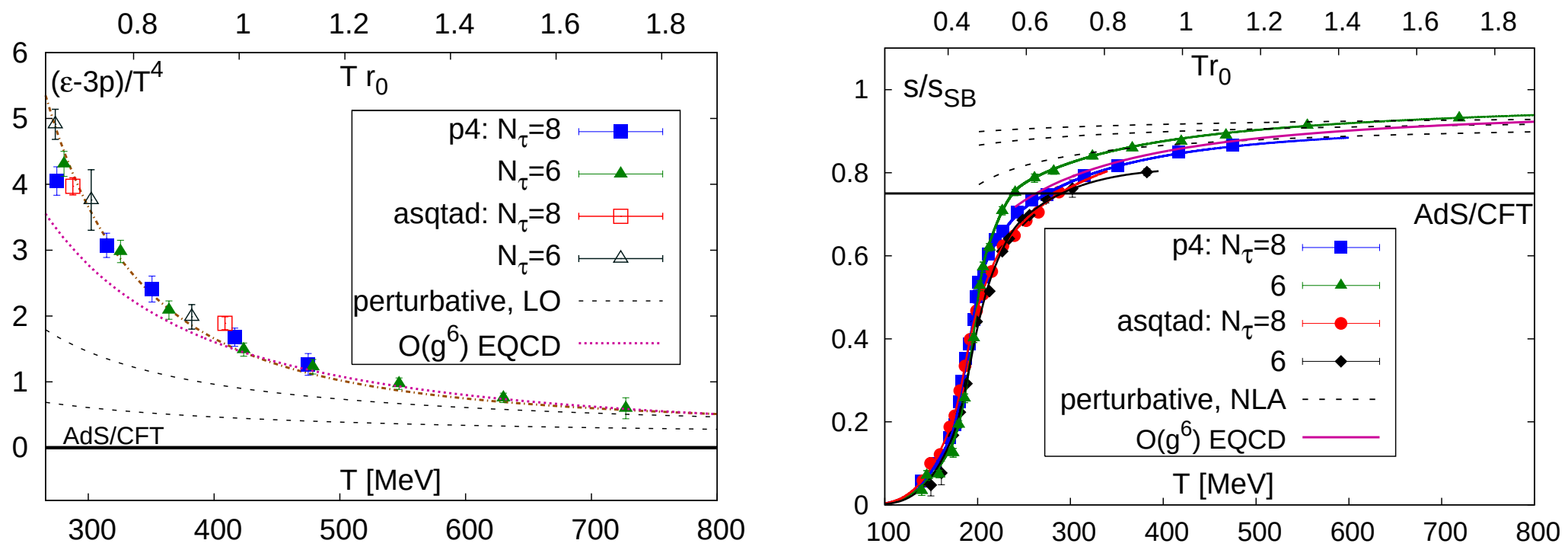

Figure 3: The trace anomaly in the high temperature region compared with weak coupling calculations and phenomenological fit (left) and the entropy density (right) normalized to the ideal gas value ($s_{SB}$) and compared with perturbation theory and EQCD. For the trace anomaly the leading order (LO) approximation is shown, while for the entropy density the resummed next-to-leading approximation (NLA) is shown. Lattice results are from Refs. [6, 7, 8, 29].

## 3. Fluctuations of conserved charges

The pressure at non-zero chemical potentials can be evaluated using Taylor expansion. The Taylor expansion can be set up in terms of quark chemical potentials or in terms of chemical potentials corresponding to baryon number $B$, electric charge $Q$ and strangeness $S$ of hadrons. The expansion coefficients in quark chemical potential $\chi^{jkl}_{uds}$ and the hadronic ones $\chi^{jkl}_{BQS}$ are related to each other. For a more detailed discussion on this see Ref. [31]. Taylor expansion can be used to study the physics at non-zero baryon density. However, the expansion coefficients also give important information about fluctuation of conserved charges, which turn out to be sensitive probes of deconfinement and chiral symmetry restoration. Fluctuations of conserved charges have been studied in detail using *asqtad* and *p*4 actions [2, 31, 8]. In Fig. 4 I show the quadratic and quartic fluctuations of $B$, $Q$ and $S$. Fluctuations are suppressed at low temperatures because conserved charges are carried by massive hadrons, but rapidly grow in the transition region $T = (185-195)$ MeV as consequence of deconfinement. At temperatures $T > 300$ MeV fluctuations are reasonably well described by ideal gas of quarks. Quartic fluctuations exhibit a peak near the transition region, which at sufficiently small quark masses can be related to the chiral transition and the corresponding critical exponents (see discussion in Refs. [8, 32]). As one can see from the figure quadratic fluctuations show visible cutoff dependence at low temperatures, while the cutoff dependence is small for $T > 250$ MeV.

The transition from hadronic to quark degrees of freedom can be particularly well seen in the temperature dependence of the Kurtosis, which is the ratio of quartic to quadratic fluctuations. This quantity can be measured experimentally, at least in principle. If Fig. 5 I show the Kurtosis of the baryon number as a function of the temperature. At low temperatures it is temperature independent and is close to unity in agreement with the prediction of the hadron resonance gas model. In the transition region it sharply drops from the hadron resonance gas value to the value corresponding to an ideal gas of quarks. Since the fluctuations of conserved charges are so well described by ideal quark gas it is interesting to compare the lattice results for quark number fluctuations with resummed perturbation theory. Quadratic quark number fluctuations are also called quark number susceptibilities and have been calculated in resummed perturbation theory [33, 34]. The comparison of the resummed perturbative results with lattice calculations performed with $p4$ action [29] for the quark number susceptibility divided by the ideal gas value is shown in Fig. 5. The figure shows a very good agreement between the lattice calculations and resummed perturbative results. It is also interesting to compare findings of the lattice calculations with the results obtained in strongly coupled gauge theories using AdS/CFT correspondence. In $N = 4$ SUSY there is a conserved $R$-charge. Fluctuations of the conserved $R$-charge in the strong coupling limit as well as for the non-interacting case have been calculated [35]. Therefore in Fig. 5 I also show the prediction of AdS/CFT, which turns out to be significantly below the lattice data. The same result is obtained if instead of $R-$charges one considers fundamental charges on a $D7$ brane coupled to gauge fields [36]. Let me finally note that the discretization errors in the lattice calculations of the fluctuations are small at high temperatures when calculated with the $p4$ action. The same is true for *asqtad* action [2, 8].

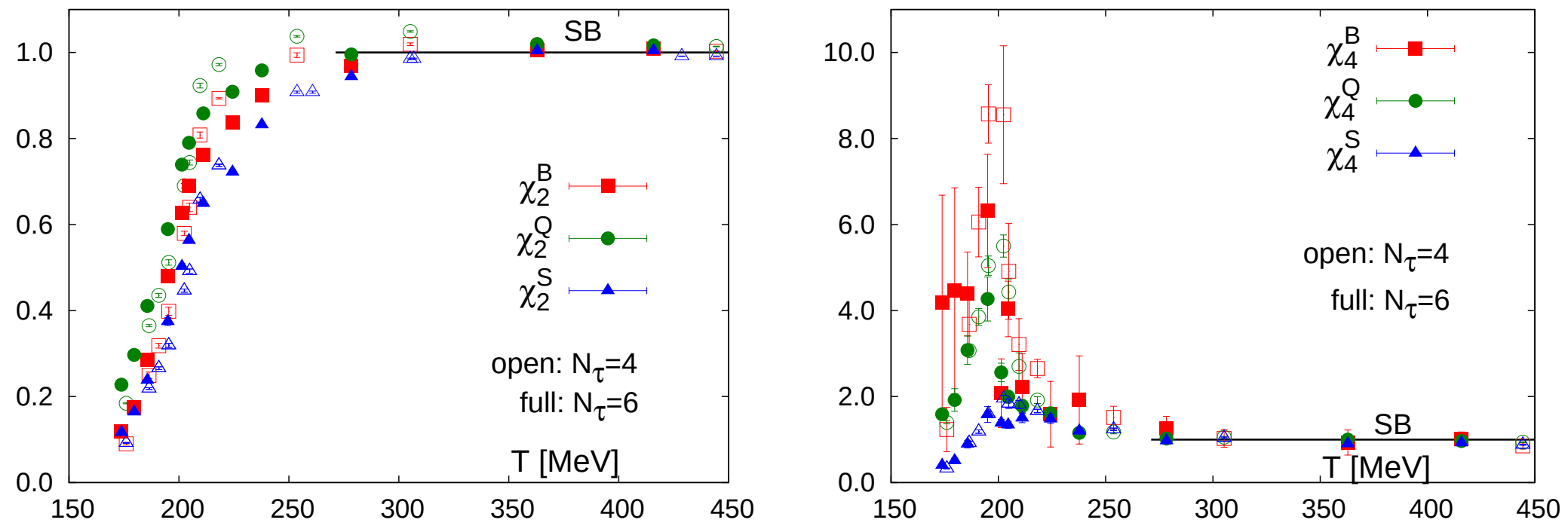

Figure 4: The quadratic (left) and quartic (right) fluctuations of $B$, $Q$ and $S$ as function of the temperature [31]. The open symbols correspond to the results obtained on $N_\tau = 4$ lattices, while the filled symbols to the ones obtained on $N_\tau = 6$ lattices.

## 4. Spatial correlation functions

Spatial correlation functions can provide additional insight into the properties of high temperature QCD. On the lattice one studies correlation functions of Wilson lines or Polyakov loops, spatial meson correlation functions and spatial Wilson loops. The correlation function of Wilson lines are related to the free energy of a quark anti-quark pair and are discussed in the contribution by Mócsy ( see also Ref. [37] for recent results). Here I would like to discuss spatial Wilson loops. It is well known that spatial Wilson loops obey an area law at all temperatures [1]

$$W(R,z) \sim \exp(-\sigma_s(T)Rz), \qquad (2)$$

[1] Strictly speaking this is only true for pure gauge theory, however, even in QCD where string breaking happens the Wilson loops show area law for all distances accessible in numerical calculations.

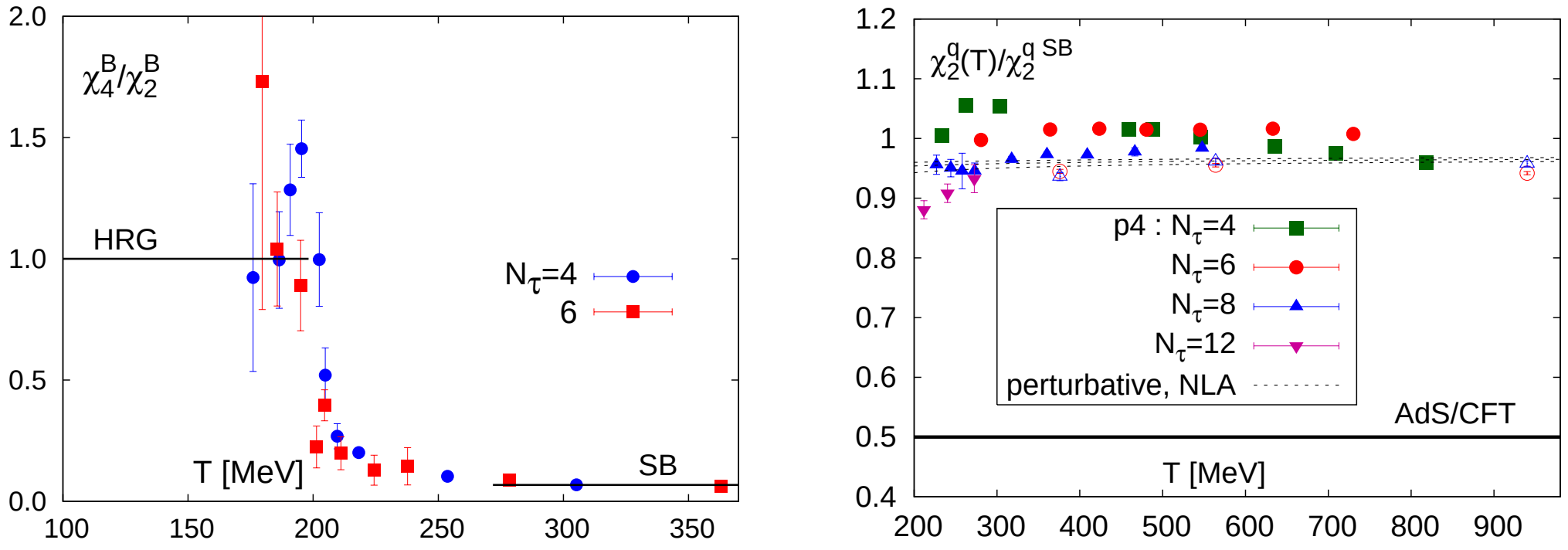

Figure 5: The Kurtosis of baryon number (left) and the quark susceptibility compared with resummed perturbative calculations (right). The open symbols on the right plot correspond to *asqtad* results for strange quark number susceptibility [2]. The resummed perturbative results were obtained in next-to-leading approximation (NLA) [34]. The numerical results for 3-flavor case have been compiled by A. Rebhan.

where $R$ is the extent of the Wilson loop in the $(x, y)$ plane and $\sigma_s(T)$ is the spatial string tension. The lattice results on the spatial string tension in $(2+1)$ flavor QCD calculated with $p4$ action for $N_\tau = 4,\ 6$ and 8 [38] are shown in Fig. 6. At low temperatures, the spatial string tension is approximately temperature independent and coincides with the zero temperature string tension. At sufficiently high temperature, however, it is proportional to $T^2$. This behavior can be understood in terms of EQCD. This effective field theory has a dimensionful gauge coupling constant $g_3^2$ and is confining [25]. Therefore one expects a non-vanishing string tension in this theory, which because of dimensional reason is $\sigma_s = c g_3^4$. The coefficient of proportionality, $c$, is know from 3-dimensional lattice calculations (see discussion in Ref. [38]). The 3d gauge coupling $g_3^2$ can be related to the gauge coupling of QCD $g(\mu)$ and this relation is known to 2-loop [39]. Using the non-perturbatively determined $c$ and the 2-loop expression for $g_3^2$ in terms of $g(\mu)$ one can calculate the prediction of EQCD for the spatial string tension. This prediction is also shown in Fig. 6 for $T/\sqrt{\sigma_s(T)}$ and compared with the lattice data. As one can see from the figure the effective theory can describe the temperature dependence of the spatial string tension down to temperature $T \simeq 300$MeV or maybe even lower. Thus the temperature dependence of the spatial string tension is perturbative and non-perturbative effects are caused solely by soft chromo-magnetic fields. Note that in the conformal strongly coupled gauge theory $T/\sqrt{\sigma_s(T)} \sim 1/\lambda^{1/4}$ (see e.g. Ref. [40] and references therein) is temperature independent and vanishes in the limit of infinite t'Hooft coupling, $\lambda \to \infty$. This is in sharp contrast with lattice calculations. To deal with this problem one can modify the AdS metric to break the conformal invariance and get a QCD-like theory. This approach is known as AdS/QCD. Attempts to model the temperature dependence of the spatial string tension using AdS/QCD were discussed in Ref. [41, 42].

Spatial meson correlation functions have also been studied on the lattice. One expects that at large distances the spatial meson correlators should decay exponentially. The screening mass that governs the exponential decay should be $\simeq 2\pi T$ at high temperatures. Lattice calculations indicate that this is the case [43], however, a quantitative comparison with EQCD has not been performed yet.

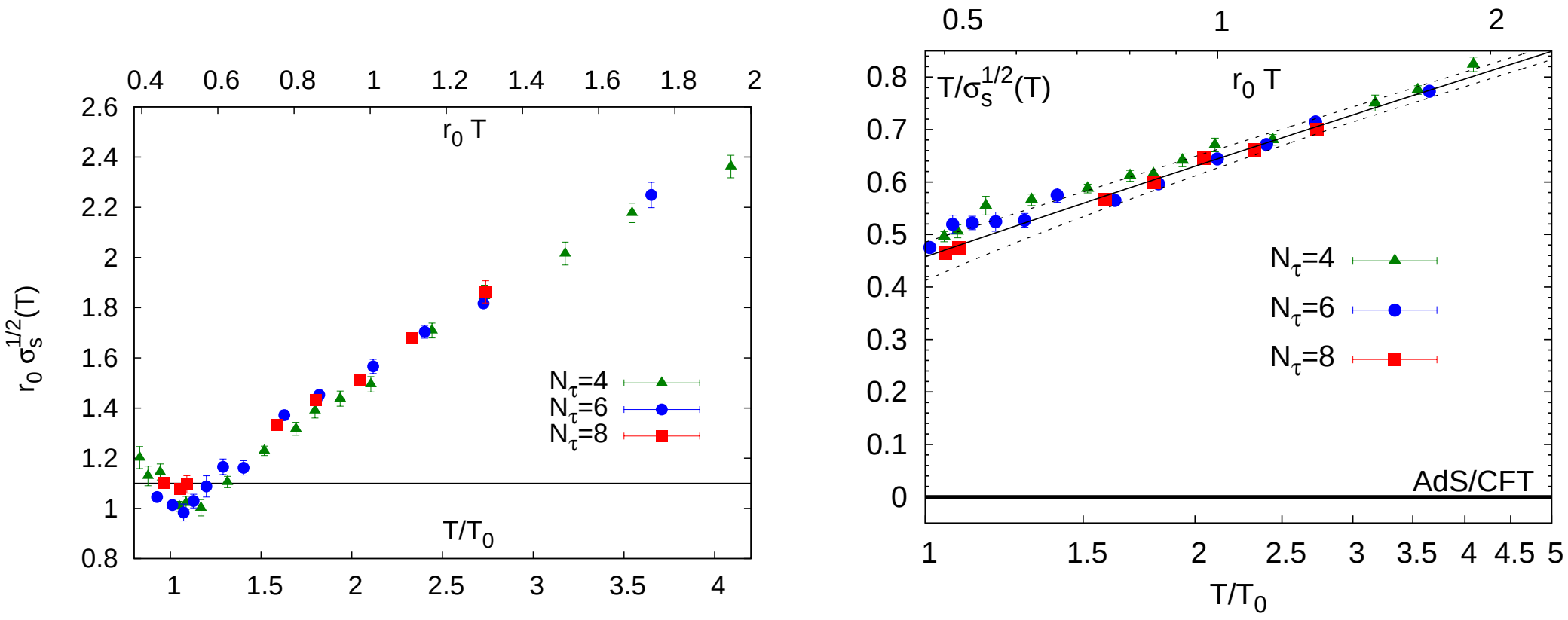

Figure 6: The square root of the spatial string tension in units of Sommer scale $r_0 = 0.469$fm (left) and the comparison of the lattice results for $T/\sqrt{\sigma_s(T)}$ with EQCD (right).Here $T_0 = 200$MeV. The dashed lines show the uncertainties in the predictions of EQCD (see Ref. [38] for details).

## 5. Conclusion and outlook

In this contribution I have discussed lattice calculations of the equation of state, fluctuations of conserved charges and spatial correlation functions with improved *p*4 and *asqtad* actions. We have seen that energy and entropy densities show a rapid rise in the temperature interval $T = (185 - 195)$ MeV indicating deconfinement. The deconfinement is also seen in the rapid rise of the quadratic fluctuations of the conserved charges in the same temperature interval. At high temperatures ( $T > 300$ MeV ) equation of state and fluctuations of conserved charge can be understood using weak coupling calculations. This is also supported by a study of spatial correlation functions. The temperature region $T < 300$ MeV which is most relevant for RHIC in general is not accessible to weak coupling techniques. I would like to stress that in the high temperature region cutoff effects are under control if improved fermion actions are used. This is not the case in the transition region and at low temperatures. Here we see significant cut-off dependence in thermodynamic quantities which may explain the disagreement between the hadron resonance gas model and lattice. Possible large cutoff effects at low temperatures may also explain the big discrepancy in quantities like, chiral condensate, strangeness fluctuation and renormalized Polyakov loop obtained with *p*4 as well as *asqtad* actions and *stout* action [9, 44]. Calculations with *stout* action indicate a shift of the transition region by 20 – 30 MeV compared to calculations with *p*4 and *asqtad* actions [9, 44]. Though the stout action does not improve the quark dispersion relation it has smaller $(a\Lambda_{QCD})^2$ discretization errors from flavor symmetry breaking effects present in the staggered fermion formulation. This is presumably the reason why cutoff dependence of the chiral condensate and strangeness fluctuations is smaller for the *stout* action at low temperatures. Hopefully the ongoing calculations of HotQCD collaborations on finer lattices corresponding temporal extent $N_\tau = 12$ as well as using highly improved staggered fermion action will shed some light on this problem.

## Acknowledgments

This work was supported by U.S. Department of Energy under Contract No. DE-AC02-98CH10886. I would like to thank Z. Fodor, F. Karsch, S. Katz, M. Laine, K. Rummukainen and D. Teaney for useful discussions and correspondence. I would like to thank A. Rebhan for sharing his numerical results on the resummed perturbative calculations in 3-flavor QCD and correspondence. Finally I would like to thank J. Casalderrey Solana for correspondence on the spatial string tension calculations in AdS/CFT and D. Teaney for the discussion of fluctuations in super-symmetric gauge theories. I would like to thank my colleagues in RBC-Bielefeld and HotQCD collaborations for fruitful joint work.